# Tuning Ferroelectrics to Antiferroelectrics in Multiferroic $La_xSr_{1-x}Fe_{12}O_{19}$ Ceramics

Cong-Cong Duan, Guo-Long Tan*

*State Key Laboratory of Advanced Technology for Materials Synthesis and Processing,
Wuhan University of Technology, Wuhan 430070, China*

**Abstract**

The combination of antiferroelectricity (AFE) and ferromagnetism (FM) in one structure would allow the development of new type of multiferroic candidates, which be applicable not only in magnetoelectric memories but also in novel energy storage devices. Here we propose a novel type of multiferroic candidate $La_xSr_{1-x}Fe_{12}O_{19}$, whose room temperature state could betuned from ferroelectrics (FE) to antiferroelectrics by changing *x* from 0 to 0.5. The emphasis of this paper will be focused on the $La_{0.5}Sr_{0.5}Fe_{12}O_{19}$ system, in which full AFE and FM coexist. The pure antiferroelectric behavior in $La_{0.5}Sr_{0.5}Fe_{12}O_{19}$ ceramics is demonstrated by double polarization-electric field (P-E) hysteresis loops, which are fully separated by a linear antiferroelectric AFE component with zero net polarization. The material of $La_{0.2}Sr_{0.7}Fe_{12}O_{19}$ with the intermediate composition exhibits a hybrid ferroelectric/antiferroelectric state. The recoverable energy density of the antiferroelectric $La_{0.5}Sr_{0.5}Fe_{12}O_{19}$ phase reaches $U_{re}$~14.3 $J/cm^3$. This material demonstrates strong magnetoelectric coupling and giant magnetoresistance (GMR) effect. A 1.1T magnetic field generates electronic polarization up to $0.95\mu C/cm^2$, reduce the resistance by 117%, enhances dielectric constants by 540% and right shifts the maximum dielectric loss peak by 208 kHz. The combined functional responses provide an opportunity to develop novel multifunctional electric and energy storage devices.

**Keywords:** Antiferroelectrics, Energy Storage, Magnetism, Magnetoelectric Coupling.

## 1. Introduction

Multiferroic materials can be used to realize a new generation of multifunctional devices, such as new magnetoelectric (ME) sensing devices, spintronic devices, high performance information storage devices. [1] For multiferroic material with ME coupling, a voltage rather than a current is used to regulate the magnetization direction and minimizes joule heat dissipation. Their use can fundamentally solve the problem of high energy consumption, and a new generation of memories with ultra-low power consumption can be realized, fast information storage and high processing speed. In addition, it is important to understand and

* Corresponding author; Tel: +86-27-87651837; fax: +86-27-87879468. Email address: gltan@whut.edu.cn

master the interaction and order law of the new generation of multiferroic materials that combine antiferroelectrics (AFE) and ferromagnetism (FM), which would make it possible to observe new quantum phenomena.[2] More importantly, the control of these new antiferroelectric phenomena in multiferroic materials would provide an opportunity to develop new generation of energy storage devices that can store both electrical and magnetic energy.[3, 4] The ME coupling in multiferroic materials can induce electric polarization under a magnetic field. The ME polarization generates a spin current or voltage, which can provide supplementary energy upon the electric filed induced charges and thus enhance the efficiency of energy storage in multiferroic capacitors. However, very few multiferroic materials exhibit both AFE and FM.[4, 5]

The first antiferroelectric behavior observed in a multiferroic family reportedly appeared in the $(Bi_{1-x}Re_x)FeO_3$ (Re=La, Sm) thin film system, which exhibits a hybrid state of ferroelectrics (FE) and antiferroelectrics (AFE),[6] whereas full multiferroic AFE feature was achieved in the $Dy_{0.75}Gd_{0.25}FeO_3$ system, which exhibits pure double *P-E* loops at 1.85K.[7] The rare earth orthoferrites $RFeO_3$ exhibit a combination of AFE and antiferromagnetism, but the antiferroelectric state appears only at very low temperature (1.85K) and low saturation polarization (0.15μC/cm$^{-1}$).[7] The discovery of a new type of multiferroic candidate with both AFE and FM at room temperature (RT) would be scientifically significant, because it may not only be useful for multiferroic memories, which could in principle enable data to be written electrically and read magnetically, but also could extend the application of multiferroic materials to the field of energy storage devices that could be electrically and magnetically charged simultaneously.

Conventional antiferroelectric materials are used to store electric energy, they play a key role in mobile electronic devices, stationary power systems, and hybrid electric vehicles.[8, 9] They include mainly non-magnetic perovskite compounds, such as $PbZrO_3$ [10], $AgNbO_3$ [11], $(Bi,Na)TiO_3$[12], $(Pb,La)(Zr,Sn,Ti)O_3$,[13] and others.[14] Antiferroelectric materials have higher energy storage density than their ferroelectric (FE) counterparts because of their double hysteresis loops and much lower remnant polarization.[15] These perovskite capacitors, however, can be charged only by using a one-dimensional electric field. Multiferroic capacitors exhibiting both AFE and FM would enable the storage of not only the electrical energy but also magnetic energy. This capability would make it possible to charge capacitors using a magnetic field to obtain higher energy storage density. M-type rare earth hexaferrites, which exhibit both AFE and FM, may be ideal candidates for testing this idea. M-type hexaferrites have been widely used in permanent magnet bodies, gigahertz-band microwave devices, high-density memory, etc, because of their strong magnetism. They were recently reported to be ferroelectrics with a large electric polarization at RT.[16, 17, 18] $SrFe_{12}O_{19}$ undergoes a phase transition from ferroelectrics to antiferroelectrics at approximately 235°C.[18] In our recent work,[19] we found that the replacement of 20% of the $Sr^{2+}$ ions in $SrFe_{12}O_{19}$ with $La^{3+}$ ions shifted this critical temperature to near RT, where the new compound enters a hybrid ferroelectric/antiferroelectric state. Motivated by this consideration, we then conceived a new type of multiferroic candidate that integrates full AFE and strong FM in a single phase by increasing the concentration of substituted La ions up to 50%. We demonstrate how the state of $SrFe_{12}O_{19}$ is gradually transformed from ferroelectrics to antiferroelectrics with increasing La ion content. The heavier substitution generates fruitful new physical phenomena, such as GMR effect and very large magnetic dielectric (MD) response. We discuss potential application of these phenomena in energy storage devices.

## 2. Experimental Procedure:

To tune M-type strontium hexaferrite ($SrFe_{12}O_{19}$) from ferroelectrics to full antiferroelectrics, we replaced (1-$x$) percent of the $Sr^{2+}$ ions in $SrFe_{12}O_{19}$ with $x$ $La^{2+}$ ions to obtain $La_xSr_{(1-x)}Fe_{12}O_{19}$ ($0 \leq x \leq 0.5$). The synthesis method of $SrFe_{12}O_{19}$ multiferroic ceramics is described in the literature,[18] here we present the fabrication procedure of $La_{0.5}Sr_{0.5}Fe_{12}O_{19}$ ceramics. A polymer precursor method was applied to prepare $La_{0.5}Sr_{0.5}Fe_{12}O_{19}$ nanometer powders first. Lanthanum acetate [$La(CH_3COO)_2·3H_2O$] (99.9%, Aladdin), strontium acetate [$Sr(CH_3COO)_2·3H_2O$] (99.9%, Aladdin), and ferric acetylacetonate [$C_{15}H_{21}FeO_6$] (99.9%, Alfa Aesar) were used as lanthanum, strontium, and iron sources, respectively. First, 6.0325 g of iron acetylacetone was dissolved in a mixture solvent of 50 mL of ethanol and 70 mL of acetone in a 250 mL three-flask inside a glove box. The mixture solution was heated and stirred on a mantle at 70°C for 6 h. Outside the glove box, lanthanum acetate and strontium acetate were dissolved in glycerin in a 50 mL flask at a (La+Sr)/Fe atomic ratio of 1:10, where the Sr/La ratio was set to 0.5:0.5. The dispersion solution was distilled at 120°C for 2 hours in a rotary evaporator until the lanthanum and strontium acetates were fully dissolved. Then the lanthanum and strontium precursors were transferred to the glove box, where the iron precursor and La+Sr precursor were mixed and held on the mantle at 70°C for another 8 h. Next, 60 mL of an ammonia solution was added into the mixture solution, which was further heated on the mantle at 70°C for more than 48 h to precipitate the La+Sr and Fe ions completely. Finally, the particles suspended in the solution were separated from the solvent by centrifugation and then calcined into powders at 450°C for 1 h. After 30 min of grinding, the powder was calcined a second time at 800°C for 1 hour to remove organic impurities. Pure $La_{0.5}Sr_{0.5}Fe_{12}O_{19}$ powder was thus obtained. The powder was pressed into pellets with a diameter of 6.2 mm and a thickness of ~0.55mm. The pellet specimens were sintered into a ceramic at 1350°C for 2 h and subsequently annealed three times in oxygen at 800°C, where each annealing period lasted 3 h, and the surface was oriented upward and downward. Afterwards, both side surfaces of the ceramic specimen were coated with silver electrodes, which were solidified at 820°C for 15 min. The structure of the specimens was determined by X-ray diffraction (XRD) measurement using Cu Ka radiation, and the magnetic properties were measured by a Quantum Design physical property measurement system (PPMS). The polarization–electric field ($P–E$) hysteresis loop was measured using a ferroelectric measurement system based on a Sawyer–Tower circuit. The temperature-dependent dielectric spectrum and impedance spectrum were measured using a Hioki IM 5330 LCR meter. The ME polarization performance and MD response were measured using a Keithley 2450 source meter and a Microtest Precision 6630 LCR meter in a frequency range of 10 Hz to 10 MHz by applying a variable DC magnetic field.

## 3. Results and Discussion

### 3.1 Structure and Chemical Identification of $La_{0.5}Sr_{0.5}Fe_{12}O_{19}$ System

Because the multiferroic behavior of $SrFe_{12}O_{19}$ (denoted as L0, $x=0$) and $La_{0.2}Sr_{0.7}Fe_{12}O_{19}$ (denoted as L2, $x=0.2$) has been investigated in detail elsewhere,[18, 19] this study focused on $La_{0.5}Sr_{0.5}Fe_{12}O_{19}$ (denoted as L5, $x=0.5$) system. The structure of the $La_{0.5}Sr_{0.5}Fe_{12}O_{19}$ ceramic powder was found to be the same as that of $SrFe_{12}O_{19}$ on the basis of XRD pattern, as shown in Figure 1. The red lines show the standard diffraction spectrum of $SrFe_{12}O_{19}$ (PDF#33-1340). The diffraction peaks of $La_{0.5}Sr_{0.5}Fe_{12}O_{19}$ are in good agreement overall with those of the standard, suggesting that the replacement of the 50% Sr in $SrFe_{12}O_{19}$ with equal La did not change its crystal structure, instead, a pure solid solution of $La_{0.5}Sr_{0.5}Fe_{12}O_{19}$ was successfully obtained.

Figure 2 displays XPS La 3d spectrum of $La_{0.5}Sr_{0.5}Fe_{12}O_{19}$, that of $LaFeO_3$ is also shown for comparison. The La 3d5/2 and 3d3/2 lines of $La_{0.5}Sr_{0.5}Fe_{12}O_{19}$ appear at 833.3eV and 849.7eV [Figure 2A (a)], whereas those of $LaFeO_3$ appear at 834.6eV and 850.8eV respectively [Figure 2A (b)]. La ions with a higher valence reportedly shifted La 3d bands to higher binding energy in XPS measurement of metallic La exposing to air. [20, 21] The longer is the exposure time of metallic La to air, the higher is the valence charge of La ions, and the larger is the binding energy of La 3d lines. [20] There is a chemical shift of around 1.1~1.3eV for The La 3d lines of $La_{0.5}Sr_{0.5}Fe_{12}O_{19}$ were chemically shifted to lower binding energy by approximately 1.1–1.3 eV compared those of $LaFeO_3$, indicating that the valence charge of the La ions in $La_{0.5}Sr_{0.5}Fe_{12}O_{19}$ is lower than that in $LaFeO_3$. Therefore, the La ions in $La_{0.5}Sr_{0.5}Fe_{12}O_{19}$ are bivalent ($La^{2+}$), whereas those in $LaFeO_3$ are trivalent ($La^{3+}$).

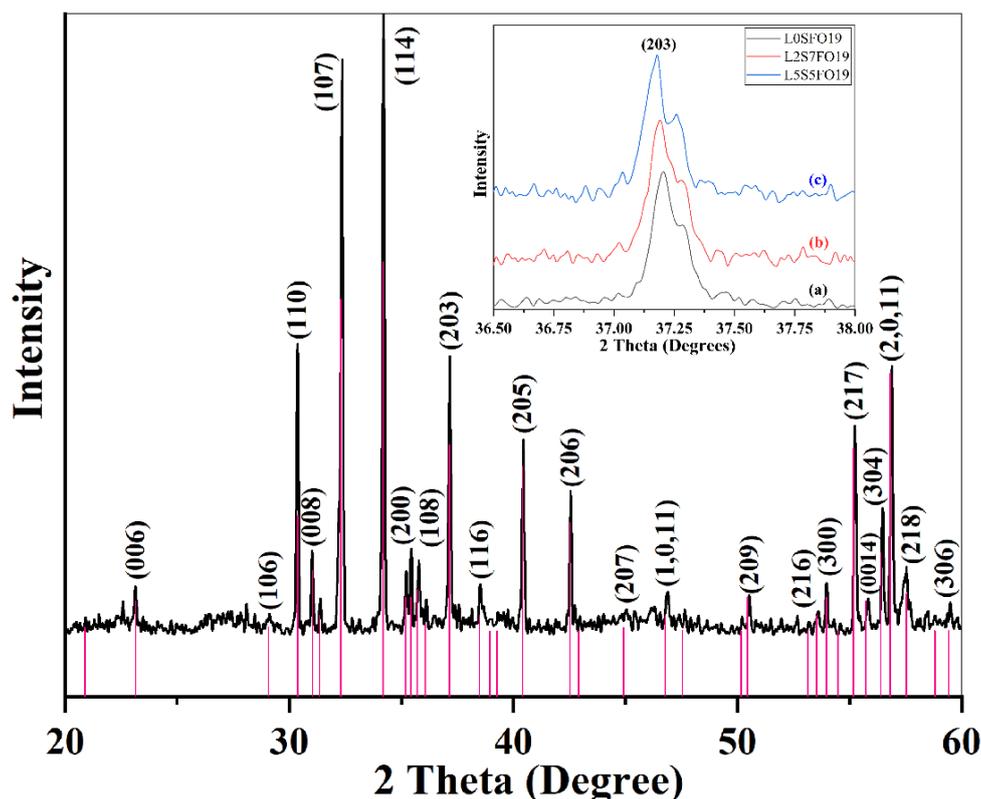

Figure 1: *XRD pattern of $La_{0.5}Sr_{0.5}Fe_{12}O_{19}$ sintered at 1350℃ for 3 h;* standard diffraction spectrum of $SrFe_{12}O_{19}$ (PDF #33-1340) is shown below. *Inset: Diffraction peak of (203) lattice plane of (a) $SrFe_{12}O_{19}$, (b) $La_{0.2}Sr_{0.7}Fe_{12}O_{19}$, and (c) $La_{0.5}Sr_{0.5}Fe_{12}O_{19}$.*

The Raman bands (Figure S1 in the Supplementary Materials) of $La_{0.5}Sr_{0.5}Fe_{12}O_{19}$ are in good agreement with those of pure $SrFe_{12}O_{19}$, and there are no additional bands indicating other oxide impurities, which also suggests that we have successfully replaced 50% of the Sr ions with La ions and obtained a solid solution of $La_{0.5}Sr_{0.5}Fe_{12}O_{19}$. According to this formula, La ions would be bivalent ($La^{2+}$) instead of trivalent, being in good agreement with the XPS valence spectrum (Figure S2 in the Supplementary Materials). Consequently, the valence charge of 0.5La+0.5Sr would be +2, whereas 12 $Fe^{3+}$ ions provide +36 positive charges, the total positive charge is thus +38, which may balance the total negative charge (-38) of 19 $O^{2-}$ ions in the $La_{0.5}Sr_{0.5}Fe_{12}O_{19}$ system. The charge balance in this formula would leave no vacancies in the lattice.

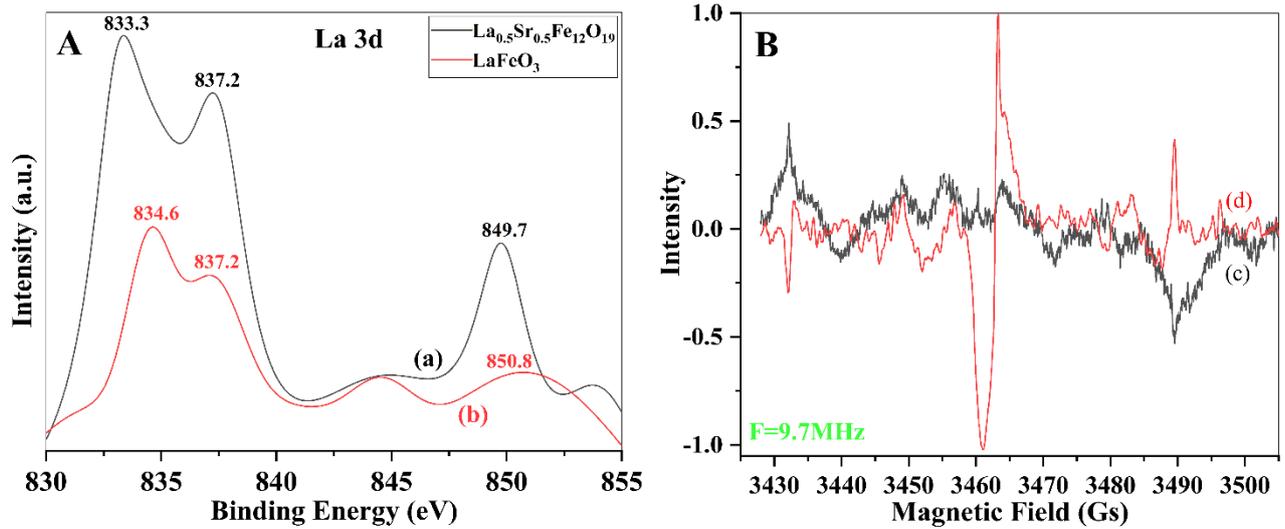

Figure 2-A: *XPS La 3d lines of $La_{0.5}Sr_{0.5}Fe_{12}O_{19}$ (c) and $LaFeO_3$ (b) systems; -B: Electron Spin Resonance (ERS) spectrum of $La_{0.5}Sr_{0.5}Fe_{12}O_{19}$ ceramics, (c) being annealed in oxygen and (d) annealed in vacuum.*

The presence of oxygen vacancies was evaluated by electron spin resonance (ESR) spectroscopy. Figure 2B (c) and (d) show the ESR spectra of $La_{0.5}Sr_{0.5}Fe_{12}O_{19}$ ceramics annealed in oxygen and annealed in vacuum, respectively. The ESR spectrum (Figure 2B(c)) indicates that there are no oxygen vacancies in the $La_{0.5}Sr_{0.5}Fe_{12}O_{19}$ ceramic annealed under $O_2$, where most of the oxygen vacancies were removed during annealing. However, heat treatment of the ceramic in vacuum resulted in a high concentration of oxygen vacancies, as indicated by a strong resonance peak at 3461 G in Figure 2B (d), which results from the oscillation of unpaired electrons at the energy level of the oxygen vacancies. The spectrum of the specimen annealed in $O_2$ atmosphere does not show this resonance peak, indicating that there are no vacancies in the $O_2$-treated $La_{0.5}Sr_{0.5}Fe_{12}O_{19}$ specimen. The XPS O 1s band in Figure S5 (Supplementary Materials) confirms that no oxygen vacancies appear in the in $La_{0.5}Sr_{0.5}Fe_{12}O_{19}$ ceramic.

### 3.2 Tuning Ferroelectricity to Antiferroelectricity in $La_{1-x}Sr_xFe_{12}O_{19}$ System

The ferroelectric state of $La_xSr_{1-x}Fe_{12}O_{19}$ system can be tuned by varying $x$. $SrFe_{12}O_{19}$ ($x=0$) is ferroelectric,[18] whereas $La_{0.2}Sr_{0.7}Fe_{12}O_{19}$ ($x=0.2$) exhibits a mixture state of FE and AFE.[19] We considered the state for $x > 0.2$. By increasing the La concentration to $x=0.5$, we obtained a pure antiferroelectric state in $La_{0.5}Sr_{0.5}Fe_{12}O_{19}$ ($x=0.5$).

The transition from FE to AFE with changes in La concentration ($x$) in $La_xSr_{1-x}Fe_{12}O_{19}$ system is illustrated in Figure 3. The standard electric polarization (P-E) hysteresis loop with full saturation in Figure 3A indicates that the $SrFe_{12}O_{19}$ (L0) ceramics are in a ferroelectric state, [18] whereas the double P-E hysteresis loops mixing with intermediate void loop in Figure 3B indicate that $La_{0.25}Sr_{0.75}Fe_{12}O_{19}$ presents in a hybrid ferroelectric/antiferroelectric state. Figure 3C shows the standard double *P-E* hysteresis loops fully separated by a linear antiferroelectric component, which can be considered as the experimental evidence to demonstrate that for the $La_{0.5}Sr_{0.5}Fe_{12}O_{19}$ ceramics are in a pure antiferroelectric state.

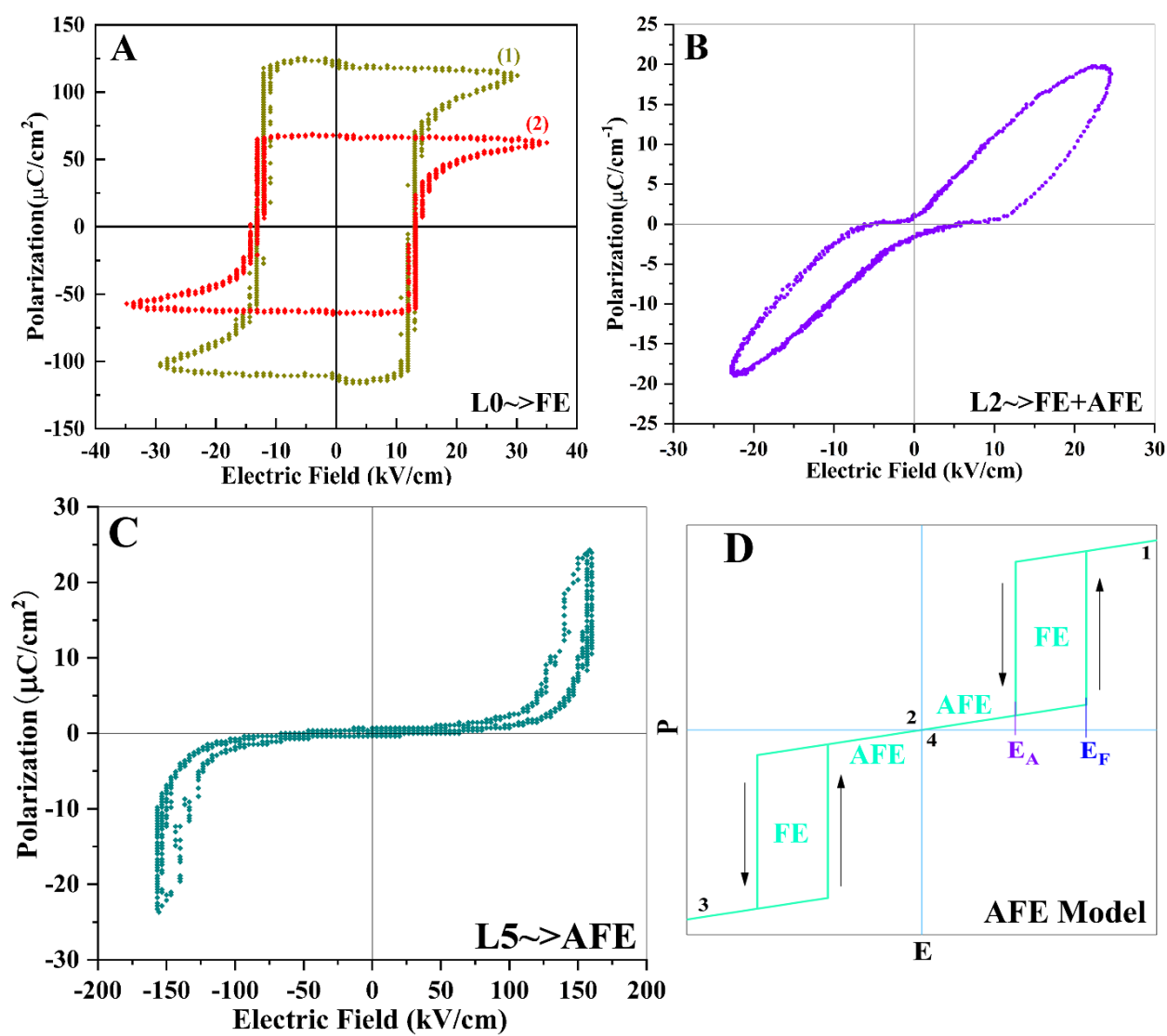

Figure 3 A: Standard ferroelectric *P-E* hysteresis loops of $SrFe_{12}O_{19}$ ceramics in full saturation, loops (1) and (2) were measured in the same specimen under different electric fields. B: *P-E* hysteresis loops of $La_{0.2}Sr_{0.7}Fe_{12}O_{19}$ ceramics, where the combination of the double *P-E* loops with an intermediate void loop suggests a hybrid ferroelectric/antiferroelectric state. C: Standard double *P-E* hysteresis loops of $La_{0.5}Sr_{0.5}Fe_{12}O_{19}$ ceramics separated by a linear

antiferroelectric component. D: Schematic diagram of double P-E hysteresis loops indicating AFE. L0, L2, and L5 represent $SrFe_{12}O_{19}$, $La_{0.2}Sr_{0.7}Fe_{12}O_{19}$, and $La_{0.5}Sr_{0.5}Fe_{12}O_{19}$, respectively. All the *P-E* hysteresis loops were measured at room temperature.

A schematic diagram of the antiferroelectric polarization switching under an electric field is displayed in Figure 3D. Two *P-E* hysteresis loops represent ferroelectric polarization (P) component on opposite sides near the saturated electric fields. The double loops are separated by a linear antiferroelectric component. In the antiferroelectric region, the positive polarization vectors have the same amplitude as the negative ones but are aligned in parallel in opposite directions and thus counteract each other; the macroscopic ***P*** vanishes to net zero and exhibits a linear relationship with E.[19] Greater linearity of the antiferroelectric component indicates a purer antiferroelectric state. Under weaker electric fields, the polarization vectors are randomly aligned, and the polarizations cancel each other. This cancellation of randomly aligned polarization in the antiferroelectric phase results in almost zero polarization on the macroscopy scale. Therefore the net polarization approaches to zero within the antiferroelectric region, where the *P–E* loop has a linear component.

Figure 3C shows that the linear component almost overlaps the x-axis, where the net polarization approaches to zero through the cancellation of oppositely aligned polarization vectors in the full antiferroelectric phase of the $La_{0.5}Sr_{0.5}Fe_{12}O_{19}$ system. The polarization is aligned along the direction of the switching electric field only when the electric field exceeds $\pm E_A$; thus, the ferroelectric component, which is indicated by double *P-E* hysteresis loops, appears only within the range of $E_A \sim E_F$. The forward switching (AFE-to-FE) field $E_A$ is $\pm 115$ kV/cm, and the backward switching (FE-to-AFE) field $E_F$ is approximately $\pm 145$ kV/cm, the switching hysteresis $\Delta E$ is approximately 30 kV/cm. The steep lines in the double *P-E* loops are almost perpendicular to the *x* axis, suggesting good saturation of the antiferroelectric loops. The maximum polarization from the double *P-E* loops is approximately 24.7 $\mu C/cm^2$ at E=159kV/cm, and the net remnant polarization is almost zero, as shown in Figure 3C. Within the antiferroelectric linear component, the polarization decreases only from 0.5 $\mu C/cm^2$ to -0.6 $\mu C/cm^2$ when the electric field changes from 50kV/cm to -66 kV/cm. The antiferroelectric polarization is ignorable compared to the ferroelectric polarization. The two switching fields ($E_A$ & $E_F$) of the $La_{0.5}Sr_{0.5}Fe_{12}O_{19}$ ceramics are comparable to those of classical antiferroelectric perovskites, such as $Pb(Zr_xTi_{1-x})O_4$ [15, 22, 23] and $AgNbO_3$,[24] which have $E_A$ values of 40 kV/cm to 170 kV/cm and $E_F$ values of 100 kV/cm to 230 kV/cm. Therefore, by changing the substitution concentration of La ions (*x*) from 0.0 to 0.5 in $La_xSr_{1-x}Fe_{12}O_{19}$ system, we are able to tune its state from ferroelectrics to antiferroelectrics while maintaining a hybrid ferroelectric/ antiferroelectric phase at an intermediate composition.

### 3.3 Dielectric Relaxation in $La_xSr_{1-x}Fe_{12}O_{19}$ Ceramics

The state tuning of the $La_xSr_{1-x}Fe_{12}O_{19}$ system at RT is revealed by changes in the phase configuration with the composition in the temperature-dependent dielectric diffusion spectrum. The real and imaginary parts of the dielectric constant usually present exhibit abnormal change at the vicinity of phase transition temperature, the state switches from ferroelectrics (FE) to antiferroelectrics (AFE) or paraelectrics (PE) across the critical temperature.

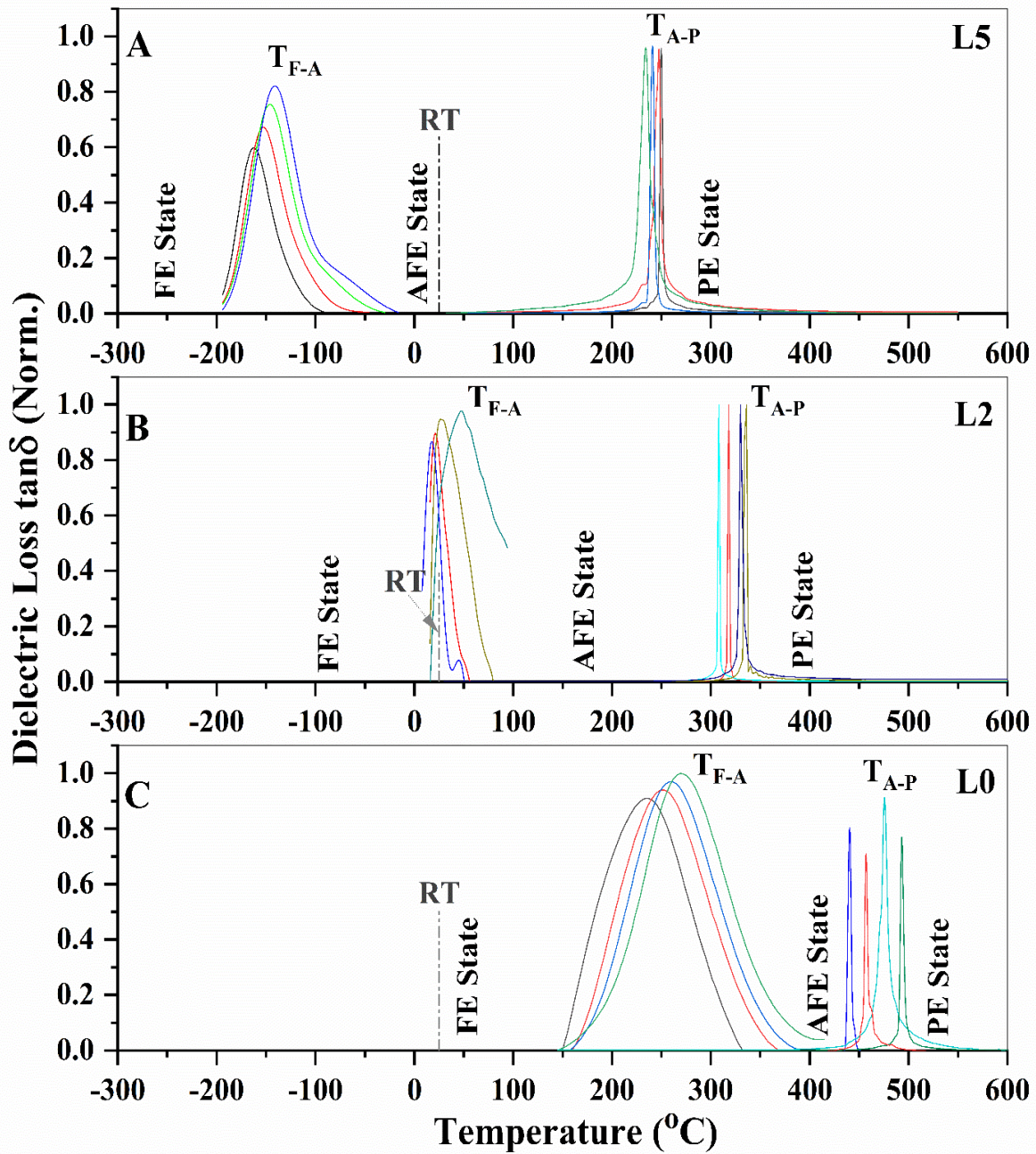

Figure 4: Temperature-dependent dielectric loss (tan δ) spectrum of $La_xSr_{1-x}Fe_{12}O_{19}$ ($x$=0, 0.2,

0.5) ceramics at frequencies of 50 kHz to 200 kHz; A: $La_{0.5}Sr_{0.5}Fe_{12}O_{19}$; B: $La_{0.2}Sr_{0.7}Fe_{12}O_{19}$, and C: $SrFe_{12}O_{19}$. L0, L2, and L5 represent for $SrFe_{12}O_{19}$, $La_{0.2}Sr_{0.7}Fe_{12}O_{19}$, and $La_{0.5}Sr_{0.5}Fe_{12}O_{19}$, respectively.

Figure 4 presents the temperature dependent dielectric loss spectrum of $La_xSr_{1-x}Fe_{12}O_{19}$ ceramics, which reveals the variation in phase configuration with $x$. The ferroelectric state switches to the antiferroelectric state near RT as the phase transition peaks shift with $x$. At RT, the L0 ($x=0$) ceramic is ferroelectric (Figure 4C), whereas the L5 ceramic ($x = 0.5$) is antiferroelectric. The L2 specimen, with an intermediate composition ($x = 0.2$), is in a hybrid ferroelectric/antiferroelectric state (Figure 4B). The ferroelectric-to-antiferroelectric phase transition peak ($T_{F-A}$) appears at 235°C (50 kHz) for the L0 ceramic, this peak downward shifts to -163°C for the L5 ceramic as $x$ changes from 0 to 0.5. In the L2 ceramic, the ferroelectric-to-antiferroelectric transition occurs at 18°C, which is approximately RT. Therefore it's not surprising that the L2 ceramic is in a hybrid ferroelectric/antiferroelectric state at RT. The critical temperature ($T_{A-P}$) for the second transition from antiferroelectric to paraelectric (PE) appears at 441°C (50 kHz) for L0, 306°C for L2 and 233°C for L5. The variation in $T_{F-A}$ and $T_{A-P}$ with $x$ is shown in Figure 4. There appears a wide temperature space of -163°C to +233°C ($T_{F-A}$ to $T_{A-P}$) to accommodate antiferroelectric state as a thermodynamically stable phase in $La_{0.5}Sr_{0.5}Fe_{12}O_{19}$ as being shown in Figure 4A (also Table S3 in the Supplementary Materials). Room temperature falls just in the middle of this scope, thus a pure antiferroelectric state appears at room temperature for L5 phase, where 0.5 $Sr^{2+}$ ions were substituted by 0.5 La ions. The clear region below the critical temperature of $T_{F-A}$ (235°C) in the L0 ceramic (Figure 4C) suggests that $SrFe_{12}O_{19}$ ceramic is ferroelectric at room temperature, which is consistent with the experimental evidence in Figure 3A and that in the literature.[18]

The temperature-dependent phase configuration of $La_xSr_{1-x}Fe_{12}O_{19}$ system is exhibited in Figure 5, which was derived from the dielectric loss datasets at 50 kHz in Figure 4. The space for hosting ferroelectric or antiferroelectric phase down shifts to lower temperature as the La concentration $x$ increases. With increasing La ion content in $La_xSr_{1-x}Fe_{12}O_{19}$, the transition temperatures $T_{F-A}$ and $T_{A-P}$ reduces, and the window for the appearance of the antiferroelectric phase becomes wider. The RT state changes from ferroelectric in the L0 ceramic to hybrid ferroelectric/antiferroelectric in the L2 ceramic, and finally switches to purely antiferroelectric in the L5 ceramic. Thus, the RT state of $La_xSr_{1-x}Fe_{12}O_{19}$ system can be tuned from ferroelectric to antiferroelectric by changing the $x$ from 0 to 0.5.

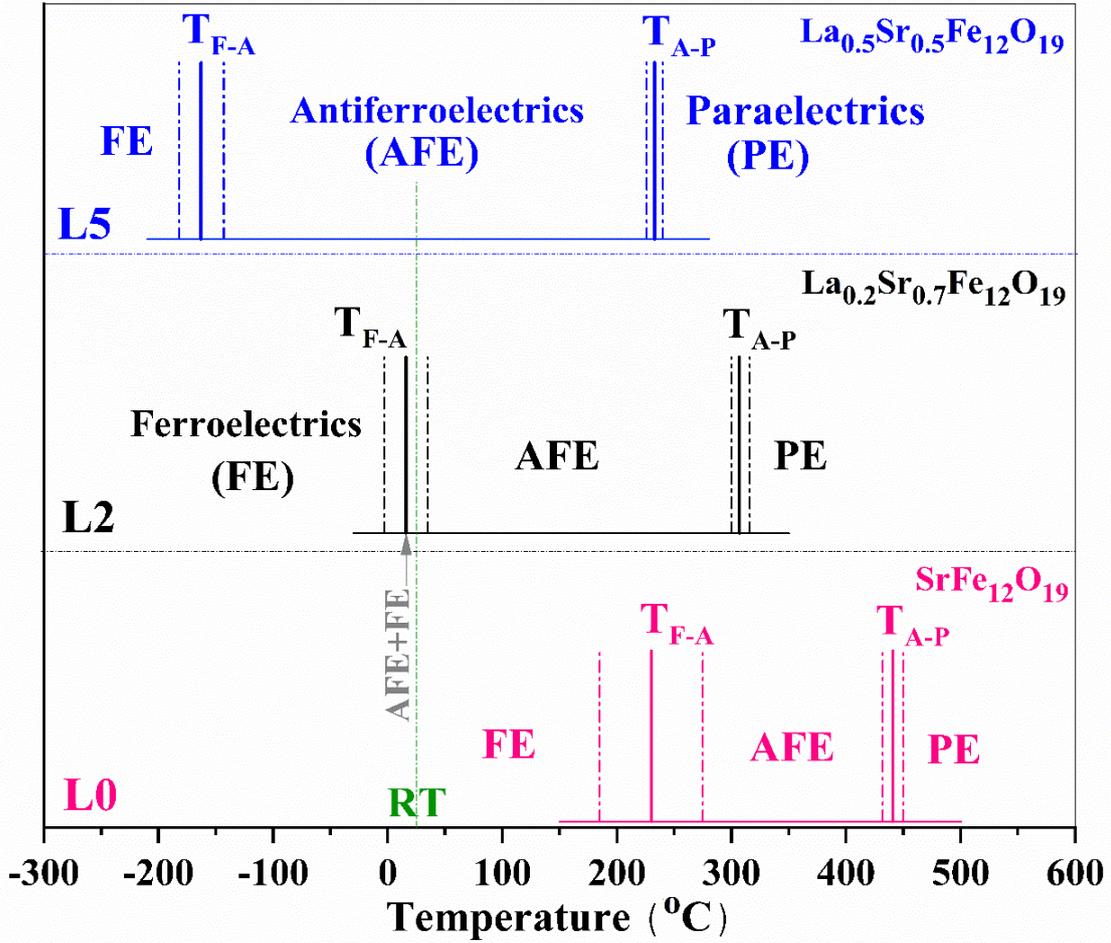

Figure 5: Diagram of the temperature-dependent phase configurations of $La_{1-x}Sr_xFe_{12}O_{19}$ system. The solid lines are peak centers, and dotted lines correspond to the full width at half maximum (FWHM). $T_{F-A}$ is the ferroelectric-to-antiferroelectric transition temperature, and the $T_{A-P}$ is the antiferroelectric-to-paraelectric transition temperature.

The spectrum of the real part of the dielectric constant reveals similar changes in phase configuration varying with temperature in the $La_xSr_{1-x}Fe_{12}O_{19}$ system. The temperature dependent ($\varepsilon_r'$) spectrum is displayed in Figure S4. The transition temperature $T_{A-P}$ of $SrFe_{12}O_{19}$ is 411ºC at 50 kHz, this peak shifts to 308ºC for the L2 ceramic and decreases further to 232ºC for the L5 ceramic. By contrast, $T_{F–A}$ shifts from 211ºC for L0 to 36ºC for L2 and -145ºC for L5. The critical temperature $T_{F–A}$ of the $La_xSr_{1-x}Fe_{12}O_{19}$ system can clearly be tuned by changing the value of $x$; this peak shifts to a very low temperature far below RT when $x$ is varied from 0 to 0.5, leaving a wide window ranging from -145ºC to 232ºC to host the antiferroelectric state for $La_{0.5}Sr_{0.5}Fe_{12}O_{19}$. Consequently, the RT state was tuned from the ferroelectric phase to the full antiferroelectric phase as $x$ was changed from 0 to 0.5, whereas the intermediate L2 phase exhibited a mixed ferroelectric/antiferroelectric state. The spectrum of the real part of the dielectric constant deduces similar outcome to the dielectric loss spectrum. However, there is retardation between the two types of spectra, the real part lags the imaginary part because of the phase difference. Therefore the transition temperatures ($T_{F-A}$ and $T_{A-P}$) in spectrum of the real part ($\varepsilon'$) are lower than those in the dielectric loss spectrum. The transition temperatures $T_{F-A}$ and $T_{A-P}$ in $La_xSr_{1-x}Fe_{12}O_{19}$ system are summarized in Table S2, and the temperature ranges in which the ferroelectric, antiferroelectric, and paraelectric states can appear are shown in Table S3.

In both the spectra of real part $\varepsilon'$ and imaginary part (tan$\delta$), the phase transition peaks exhibit strong dispersion feature, where the $T_{F-A}$ and $T_{A-P}$ peaks are highly sensitive to the frequency. In the dielectric loss spectrum of the $La_{0.5}Sr_{0.5}Fe_{12}O_{19}$ ceramics (Figure 4A), $T_{F-A}$ diffuses from -141°C to -163°C as the frequency changes from 50 kHz to 200 kHz, whereas $T_{A-P}$ decreases from 250°C to 233°C. Similar variation of $T_{F-A}$ and $T_{A-P}$ with frequency is also observed in the dielectric loss spectrum of the $La_{0.2}Sr_{0.7}Fe_{12}O_{19}$ and $SrFe_{12}O_{19}$ ceramics, as being shown in Figure 4B and C. The dispersion of the transition temperature with frequency suggests that these ceramics are relaxor ferroelectric compounds with a diffuse phase transition process. Note that the diffusion performance in the dielectric loss spectrum in Figure 4 resembles Debye relaxation rather than Maxwell-Wagner relaxation. These loss peaks are very sharp and show a much narrower relaxation temperature range than that of Maxwell-Wagner relaxation, which shows shoulder-like stunted peaks and can span a much wider dispersion range of up to 450°C because of charging at the heterogeneous interfaces between two types of materials with different dielectric constants. [25]

### 3.4 Magnetic Semiconducting Performance of $La_{0.5}Sr_{0.5}Fe_{12}O_{19}$ Compound

By definition, magnetism must be present with AFE or FE in one single phase of multiferroic materials. $SrFe_{12}O_{19}$ is a conventional magnetic compound, its magnetism didn't change greatly when 50% of the $Sr^{2+}$ ions were replaced with 0.5 $La^{2+}$ ions because the content of $Fe^{3+}$ ions, which provide unpaired electron spins for magnetism, was unchanged. The *M-H* hysteresis loops in Figure 6A reveal the magnetic performance of the $La_{0.5}Sr_{0.5}Fe_{12}O_{19}$ ceramic. A lower annealing temperature results in stronger magnetism with higher remnant magnetization because the particles are smaller. The specimen being heat treated at 1350°C holds a remnant moment of 20.3emu/g and coercive force of 4195 Oe, when the annealing temperature is reduced to 1000°C, these values change to 29.5emu/g and 3995Oe, respectively. These magnetic parameters are comparable with those of pure $SrFe_{12}O_{19}$, the replacement of

50% of the Sr in SrFe$_{12}$O$_{19}$ with equivalent La didn't change the magnetism significantly, but it produces a novel type of multiferroic candidate that combines strong FM and full AFM in one single phase at RT.

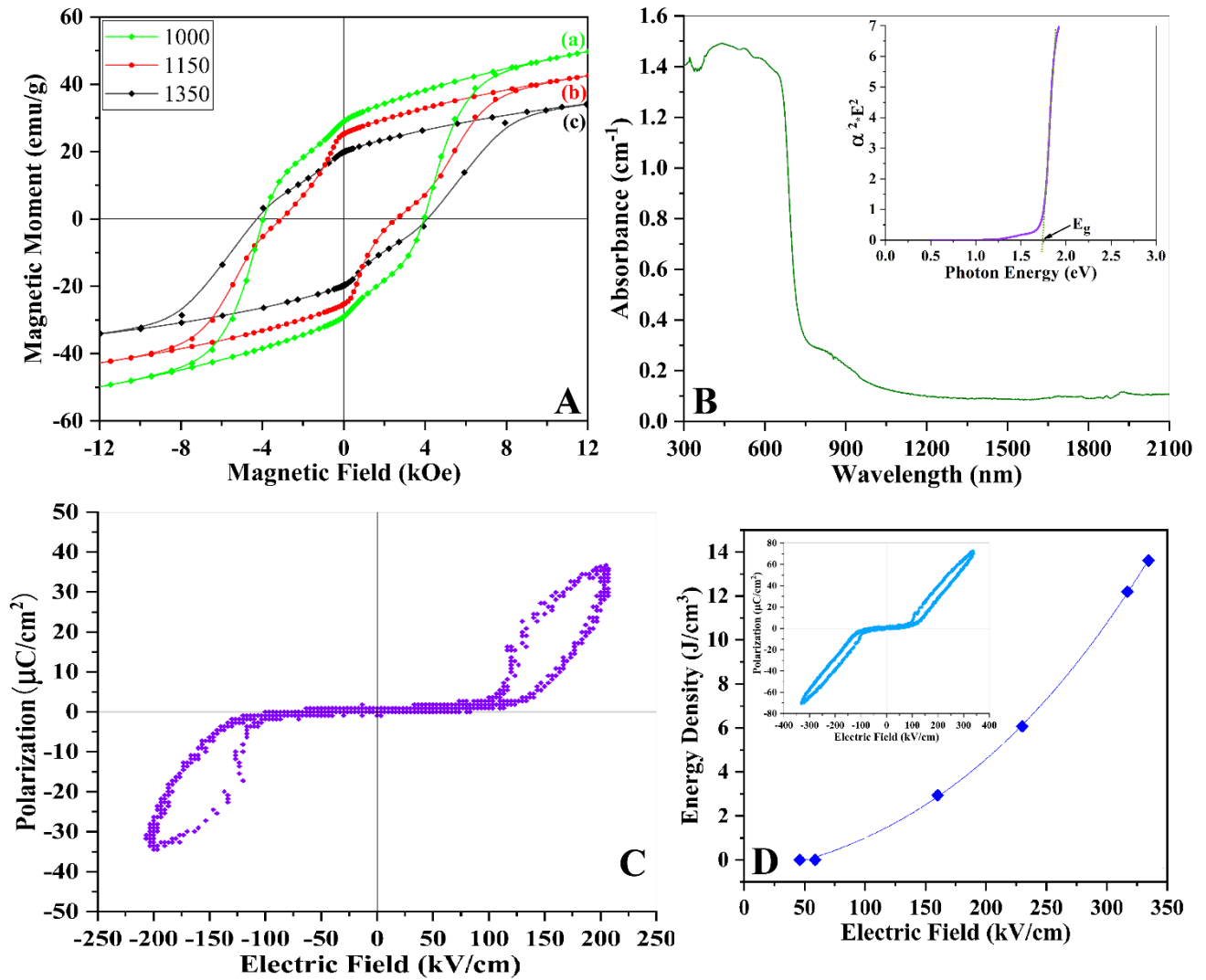

Figure 6 A: Magnetic (*M–H*) hysteresis loops of the La$_{0.5}$Sr$_{0.5}$Fe$_{12}$O$_{19}$ ceramics sintered at (a) 1000°C, (b) 1150°C and (c) 1350°C for 3. B: Optical spectrum of La$_{0.5}$Sr$_{0.5}$Fe$_{12}$O$_{19}$ ceramics; the inset shows the Tauc plot ($\alpha^2 E^2$ as a function of photon energy). C: Well-separated double P-E hysteresis loops of the La$_{0.5}$Sr$_{0.5}$Fe$_{12}$O$_{19}$ ceramics. D: Recoverable energy density of La$_{0.5}$Sr$_{0.5}$Fe$_{12}$O$_{19}$ ceramics under different electric fields. Inset shows additional double P-E hysteresis loops obtained from different La$_{0.5}$Sr$_{0.5}$Fe$_{12}$O$_{19}$ ceramics to demonstrate repeatability of the experiments.

The coexistence of AFE and FM in the La$_{0.5}$Sr$_{0.5}$Fe$_{12}$O$_{19}$ system is further revealed by the well-separated double *P-E* hysteresis loops with a linear component (Figure 6C), which gives a recoverable energy density of approximately 12 J/cm$^3$. The highest recoverable energy density in the $La_{0.5}Sr_{0.5}Fe_{12}O_{19}$ ceramics is estimated to be around 14.3 J/cm$^3$ at 320kV/cm from the hysteresis loops in the inset of Figure 6D. These antiferroelectric double P-E loops were measured from different $La_{0.5}Sr_{0.5}Fe_{12}O_{19}$ ceramic specimen, so as to demonstrate the reliability and repeatability of the experimental results. The energy density, 14.3 J/cm$^3$, is twice that of lead lanthanum zirconate titanate (PLZT) perovskite capacitors (6.7J/cm$^3$) and is the largest among those of the bulk antiferroelectric ceramics.[26] Similar energy densities have been obtained only in a 700 nm La-doped PbZrO$_3$ thin film system under a high field of 600 kV/cm [27] and in a 500-nm Sr-doped PbZrO$_3$ thin film under an even higher field of 900 kV/cm.[14] Further increases in the energy density of buffered thin film systems would require a much larger actuation electric field (>1500kV/cm).[26, 28] The field dependent energy density of the $La_{0.5}Sr_{0.5}Fe_{12}O_{19}$ ceramics is displayed in Figure 6D, where the energy density shows a nonlinear increase with increasing electric field. The advantage of the antiferroelectric state appearing in multiferroic materials is that it may provide an additional degree of freedom for storing magnetic energy, which would supply additional ME coupling energy through the magnetic field in addition to the conventional electric energy, and thus could enable the generation of a higher energy storage density than nonmagnetic perovskite capacitors.

In addition to the multiferroic functionality, $La_{0.5}Sr_{0.5}Fe_{12}O_{19}$ is also semiconducting, as revealed by the UV- Visible–near infrared optical spectrum in Figure 6B, which was measured using an HP Lambda 750 UV-Visible–near infrared spectrophotometer with an integrating sphere attachment. It shows displays a typical feature of direct-bandgap semiconductors, a cutoff edge appears within the wavelength range of 660nm~760nm, and the abrupt increase in absorption coefficient corresponds to the interband transition across the band gap region. An Urbach tail at 760nm~960nm is attributed to absorption by defects in the crystal. The inset of Figure 6B shows the Tauc plot ($\alpha^2E^2$ as a function of photon energy (E)); from this plot, the direct band gap energy was determined to be 1.74eV. This band gap energy is comparable in magnitude to that of silicon but slightly smaller than that of BiFeO$_3$. These experimental results suggest that $La_{0.5}Sr_{0.5}Fe_{12}O_{19}$ is not only a multiferroic compound, but also a kind of magnetic semiconductor.

### 3.5 Magnetoelectric Coupling Effect in $La_{0.5}Sr_{0.5}Fe_{12}O_{19}$ Ceramics

ME coupling is necessary for multiferroics to realize the magnetically read-out and electrically write-in operation in multiple state memories. For this purpose, we setup a ME device system to measure the variation in spin current and dielectric constants under different magnetic fields. The surfaces of a ceramic specimen were coated with silver electrodes. The specimen was clamped between two cupper plates, which were fixed in a plastic box. The box was placed between two electric magnets, and the electrodes were wired to Keithley 2450 sourcemeter or LCR meter outside the box for ME measurement. The changes in physical parameters with magnetic field were output to the instruments. We switched on the electric magnets and increased the magnetic field B continuously from 8 mT to 1.2T, maintained the B field at 1.2T for 20~50 s, and then decreased the B field from 1.2T to 8mT. This operation was repeated several times. Figure 7A displays the variation in ME spin current and resistance of the $La_{0.5}Sr_{0.5}Fe_{12}O_{19}$ ceramics under the continuously changing magnetic field. The quadratic lines within the duration of the first 50s originated from the initial pulse response of the 2450 sourcemeter. The ME spin current ($I$) and resistance ($R$) respond quickly to the changing magnetic field; the spin current increases, whereas the resistance decreases sharply with increasing $B$. When $B$ was held at 1.2T, $I$ and $R$ did not change. Both $I$ and $R$ recovered to their original values when $B$ was again increased 8 mT. The magnetic field was increased and decreased several times, and the spin current and resistance oscillated as rectangular waves in opposite directions.

The spin current waves were recorded by the 2450 Sourcemeter as the magnetic field $B$ was varied in the increment step of 25mT. The B induced electric polarization [$P(H)$] was obtained by integrating the spin current ($I$) over time ($t$). Figure S5 in the Supplementary Materials shows plots of the ME spin current versus time for different $B$ values. Figure 7B shows the integrated polarization [$P(H)$] varying with B. The magnetic field induced electric polarization [$P(H)$] decreases with B for B<200mT, above which $P(H)$ increases with increasing $B$. The maximum $P(H)$ value of 0.95μC/cm$^2$ was obtained at B=1.1T. The magnetic-field-induced electric polarization is reportedly around 0.4×10$^{-4}$μC/cm$^2$ at 1T and 2K in the $Ni_2V_2O_7$ system [29] as well as 0.08μC/cm$^2$ at 6.3T and 15K in the $Eu_{0.75}Y_{0.25}MnO_3$ system.[30] Therefore, the $P(H)$ value of the $La_{0.5}Sr_{0.5}Fe_{12}O_{19}$ ceramics has a much higher magnitude (more than 1 order higher) than those of other multiferroic systems and is comparable to the electric-field-induced electric polarization [$P(E)$] in $PbTiO_3$ ceramics.

The electric polarization [$P(H)$] was induced by the magnetic field through interaction of the two neighboring cycloid conic spins[31], $P(H) = A\sum_{i,j} e_{ij} \times (S_i \times S_j)$ in the $La_{0.5}Sr_{0.5}Fe_{12}O_{19}$ ceramics, where $e_{ij}$ denotes the unit vector connecting the spins $S_i$ and $S_j$, and $A$ is a constant which depends mainly on the exchange interaction and the spin-orbit interaction.

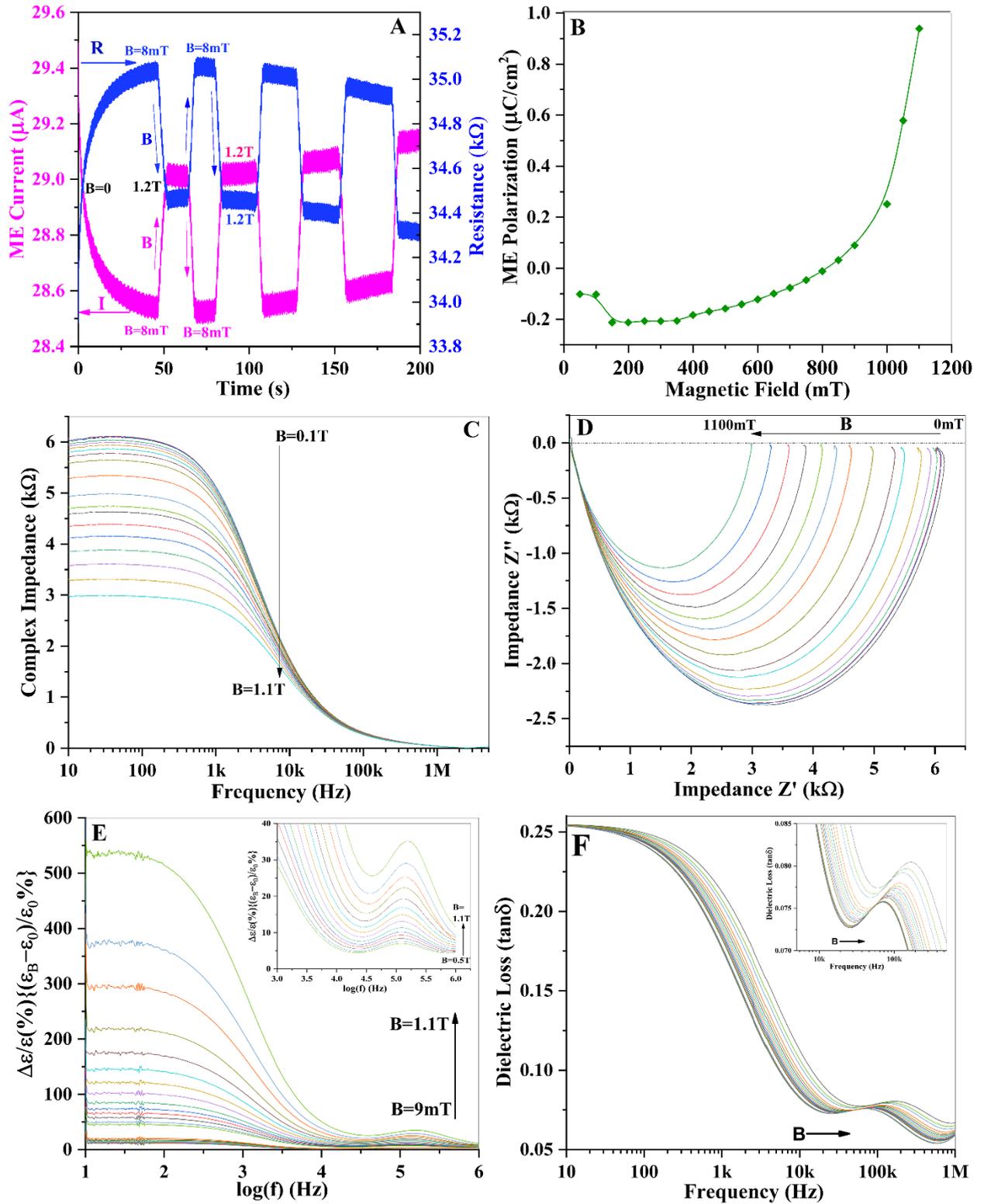

Figure 7: ME coupling of the $La_{0.5}Sr_{0.5}Fe_{12}O_{19}$ ceramics under various magnetic fields *B*. A: Variation in spin current (*I*) and resistance (*R*) with magnetic field. B: Magnetic field induced electric polarization [*P(H)*] under changing magnetic field. C: Change in impedance as a function of frequency under different B. D: Change in real and imaginary parts of impedance as a function of frequency upon varying B field. E: Change ratio in real part of dielectric

constant ($\varepsilon'$) with B field as a function of frequency. F: Variation of dielectric loss (tan$\delta$) with B field as a function of frequency. The inset shows a magnified view of the maximum dielectric loss peak (tan$\delta$) upon B field diffusing with frequency at 5 kHz-500 kHz.

$P(H)$ generates spin current as $\boldsymbol{i} = d\boldsymbol{P}/dt$. At low $B$, the directions of the conical axes in the cycloid spins of different grains are randomly distributed, some of the magnetically induced spin currents cancel each other because of the random orientations. As the strength of B field increases, an increasing number of conical spin axes are aligned along with the same direction of the magnetic field, producing a larger parallel spin current or electric polarization [$P(H)$]. Therefore a stronger magnetic field induces a larger $P(H)$.

Figure 7C shows the $B$-dependent complex impedance as a function of frequency, which was measured using a Microtest 6630 LCR meter. The impedance (Z) spectrum shows a large decrease in amplitude with increasing B field at low frequencies ($f < 1$ kHz). The module of impedance (|Z|) is 6.3k$\Omega$ at 10Hz and $B$=0 mT, and decreases gradually to 2.9k$\Omega$ as $B$ is increased to 1.1T. The maximum change in $Z$ with $B$ ($\Delta \mathbf{Z} = [\mathbf{Z(B)} - \mathbf{Z(0)})/\mathbf{Z(B)}]\%$) is -117.2%, which is comparable to the magnitude of the GMR of a Fe/Cr/Fe trilayers film system [32, 33] and of Cu/Co multilayers.[34] At high frequencies, however, this change becomes much smaller. The real part ($Z'$) and imaginary part ($Z''$) of the impedance decreases simultaneously with increasing $B$, causing the radius of the Core-Core circle to decrease with increasing $B$, as shown in Figure 7D. The real part decreases from 6.2k$\Omega$ down to 3k$\Omega$ and the imaginary part decreases from 3.2k$\Omega$ to 1.6k$\Omega$ as the applied magnetic field $B$ increases from 0T to 1.1T. The real and imaginary parts change by approximately -100%, which is also comparable to the size of the GMR in magnetic/non magnetic metal film systems.[32, 33]

Note that GMR typically appears in multiple layers film system, [35] and this is the first report of GMR in a bulk ceramic system. The origin of the GMR effect in the La$_{0.5}$Sr$_{0.5}$Fe$_{12}$O$_{19}$ ceramics may be described as follows. The spins in different grains in the ceramics are randomly aligned, and these randomly oriented spins decrease the mobility of the charge carriers (electrons and holes) through electron scattering, as they do in the Fe/Cr/Fe trilayers, [33, 35] the spin-up and spin-down currents cancel each other somehow, thus the resistance remains high when $B$=0. When a magnetic field was applied, these spins are aligned parallel to the same direction of the $B$ field. As the magnetic field becomes stronger, more spins are aligned parallel to the $B$ field. In the parallel magnetic configuration, the electrons with the same spin direction can easily pass through all the magnetic grains, and the short circuit through this channel decreases the resistance. When the spins are randomly oriented, the electrons in each channel are slowed down in every other magnetic grains, and the resistance is high.[32, 33] The uniform orientation of spins in different grains would then increase the mobility of electrons and thus reduce the resistance with increasing $B$.

The dielectric response to a magnetic field is highly consistent with the observed P(H) values, the $B$-dependent dielectric constants as a function of frequency are shown in Figure 7E. The higher B field induces larger increment of real part of dielectric constant ($\varepsilon'$) while right shifts the phase or the whole spectrum of imaginary part (tanδ) to higher frequencies. The change in $\varepsilon_r'$ with $B$ is much larger at low frequencies than at high frequencies. The maximum change in $\varepsilon_r'$ [ $\Delta\varepsilon_r'(B) = \{[\varepsilon_r'(B) - \varepsilon_r'(0)]/\varepsilon_r'(0)\} \times 100\%$ ] is approximately 540% at $B = 1.1$ T in a frequency range of 10–100Hz. The change in $\varepsilon_r'$ increases with increasing $B$. $\Delta\varepsilon_r'(B)$ decreases rapidly at frequencies below 1 kHz, at higher frequencies, the change decreases to less than 100%. The dielectric loss spectrum in Figure 7F shows the decrease in tanδ with frequency, there appears a maximum loss peak at 907 kHz at B = 0mT. Both the real and imaginary parts of the dielectric constants decrease exponentially with frequency within the range from 100 Hz to 10 kHz, this behavior is consistent the Debye relaxation model, where $\varepsilon' \& \varepsilon'' \propto 1/(1+\omega^2\tau^2)$. The entire spectrum of real part was lift up, and the maximum loss peak shifts from 97 kHz to 305 kHz as the magnetic field was increased from 0 to 1.1 T. The strong MD response is similar to that of $BaFe_{10.2}Sc_{1.8}O_{19}$ system.[36] The increase in $\varepsilon_r'$ with $B$ originates from ME coupling, which induces a spin current and thus produces additional magnetic-field-induced electric polarization P(H), which drives the increases in capacitor. The spin current increases with increasing $B$, and thus enhances P(H), since $P(H) = \int_{-\infty}^{+\infty} i\, dt$. Larger P(H) would greatly improve the dielectric constant ($\varepsilon_r'$). The relationship between the increment in $\varepsilon_r'$ and P(H) can be expressed as follows: [37]

$$\begin{aligned}\varepsilon_r(E,H) &= \frac{0.036\pi t \times P(E,H)}{U} \cdot \frac{1}{1+\omega^2\tau^2} \\ &= \frac{0.036\pi t \times [P(E)+P(H)]}{U} \cdot \frac{1}{1+\omega^2\tau^2} \\ &= \varepsilon_r(E) + \varepsilon_r(H) \quad (1)\end{aligned}$$

where $t$ is the thickness, ω is the frequency, τ is relaxation time, $U$ is the voltage, P(E) is the polarization induced by the electric field, and P(H) is the polarization driven by the magnetic field. For $B=0, \Rightarrow \varepsilon_r(E,H) = \varepsilon_r(E)$. Under a magnetic, an additional term in the expression of $\varepsilon_r(H)$ as a function of *P(H)* originating from ME coupling is added to the $\varepsilon_r(H)$ or *P(E)* term, increasing the total dielectric constant to $\varepsilon_r(E,H) = \varepsilon_r(E) + \varepsilon_r(H)$, as shown in equation 1. Higher B induces larger *P(H)* (Figure 7B), which produces higher $\varepsilon_r(E,H)$. Because of the Debye relaxation term of $1/(1+\omega^2\tau^2)$ in equation 1, the increase in $\varepsilon_r(E,H)$ is much larger at low frequencies than at high frequencies. Consequently, the frequency for the maximum loss peak ($\omega_m(H)$), which could be expressed as $\omega_m(H)\tau = \sqrt{\varepsilon_s(H)/\varepsilon_\infty(H)}$, right shifts to higher frequency because the magnitude of the increase in $\varepsilon_s(H)(\omega \to 0)$ is much larger than that of the increase in $\varepsilon_\infty(H)(\omega \to \infty)$.

## 4. Conclusion

By replacing 50% of the $Sr^{2+}$ ions in $SrFe_{12}O_{19}$ with equivalent $La^{2+}$ ions, we changed the ferroelectric state in the $SrFe_{12}O_{19}$ to antiferroelectric state in the $La_{0.5}Sr_{0.5}Fe_{12}O_{19}$ system. The ferroelectric-to-antiferroelectrics transition temperature decreased from 235°C for $SrFe_{12}O_{19}$ to -163°C for $La_{0.5}Sr_{0.5}Fe_{12}O_{19}$. Consequently, there appears a wide window of -163°C to +235°C to host the antiferroelectric L5 phase as a thermodynamically stable state, whereas room temperature falls in this scope. The pure antiferroelectric nature of the $La_{0.5}Sr_{0.5}Fe_{12}O_{19}$ system was revealed by fully separated double *P-E* hysteresis loops. The linear component separating the double *P-E* loops demonstrates zero net polarization, indicating that the polarization vectors ($\mathbf{P}\uparrow \& \mathbf{P}\downarrow$) in the antiferroelectric region are antiparallelly aligned and thus cancel each other. The highest recoverable storage energy density of the $La_{0.5}Sr_{0.5}Fe_{12}O_{19}$ system was found to be 14.3 J/cm³, which exceeds that of conventional perovskite capacitors. In addition, $La_{0.5}Sr_{0.5}Fe_{12}O_{19}$ compound also exhibits magnetic semiconducting behavior; the measured direct band gap energy is 1.74 eV, the remnant moment and magnetic coercive force were estimated to be 29.5emu/g and 3995Oe, respectively. In addition, this compound exhibited strong ME coupling and GMR. A magnetic field of 1.1T induced an electric polarization of 0.95μC/cm². The change in complex impedance at B=1.1T field was as high as -117.2%, which is comparable to the magnitude of the GMR in an Fe/Cr/Fe trilayers film system. This magnetic field lifts up the entire dielectric constant spectrum by 540% and right shifts the maximum dielectric loss peak to higher frequency by 208kHz. In summary, the $La_{0.5}Sr_{0.5}Fe_{12}O_{19}$ system is a novel type of multiferroic candidate that combines FM and pure AFE, with magnetic semiconducting behavior, strong ME coupling and high GMR.

## Acknowledgement:

The authors acknowledge the financial support from the National Natural Science Foundation of China under grant No. 11774276. The assistance of language editing service by Editage is acknowledged.

## References:

1. R. Ramesh, N. A. Spaldin, Multiferroics: progress and prospects in thin films. Nat. Mater. 6, 21-29 (2007).
2. S. W. Cheong，M. Mostovoy, Multiferroics-a magnetic twist for ferroelectricity. Nature Materials. 6, 13–20 (2007).
3. B. J. Chu, X. Zhou, K. L. Ren, B. Neese, M. R. Lin, Q. Wang, F. Bauer, Q. M. Zhang, A Dielectric polymer with high electric energy density and fast discharge speed, Science, 313, 334-336 (2006).
4. H. R. Jo, C. S. Lynch, A high energy density relaxor antiferroelectric pulsed capacitor dielectric, J. Appl. Phys. 119, 024104/1-4 (2016).
5. G. Lawes, A. B. Harris, T. Kimura, N. Rogado, R. J. Cava, A. Aharony, O. Entin-Wohlman, T. Yildirim, ,M. Kenzelmann, C. Broholm, A. P. Ramirez, Magnetically driven ferroelectric order in $Ni_3V_2O_8$, Phys. Rev. Lett. 95, 087205/1-4 (2005).
6. D. Kan, L. Palova, V. Anbusathaiah, C. J. Cheng, S. Fujino, V. Nagarajan, K. M. Rabe, I. Takeuchi, Universal behavior and electric‐field‐induced structural transition in rare‐earth‐substituted $BiFeO_3$, Adv. Funct. Mater. 20, 1108-1115 (2010).
7. Y. Tokunaga, Y. Taguchi1, T. Arima, Y. Tokura, Electric-field-induced generation and reversal of ferromagnetic moment in ferrites, Nature Physics, 8, 838-844 (2012).
8. G. R. Love, Energy-storage in ceramic dielectrics, J. Am. Ceram. Soc., 73, 323-328 (1990).
9. Z. M. Dang, J. K. Yuan, S. H. Yao, R. J. Liao, Flexible nanodielectric materials with high permittivity for power energy storage, Adv. Mater. 25, 6334-6365 (2013).
10. C. Kittel, Theory of antiferroelectric crystals, Phys. Rev. 82, 729-732 (1951).

11. H. Sakurai, S. Yamazoe, T. Wada, Ferroelectric and antiferroelectric properties of AgNbO$_3$ films fabricated on (001), (110), and (111) SrTiO$_3$ substrates by pulsed laser deposition, Appl. Phys. Lett. 97, 042901/1-4 (2010).
12. Y. Guo, Y. Liu, R. L. Withers, F. Brink, H. Chen, Large electric field-Induced strain and antiferroelectric behavior in (1-$x$)(Na$_{0.5}$Bi$_{0.5}$)TiO$_3$-$x$BaTiO$_3$ ceramics, Chem. Mater. 23, 219-228 (2011).
13. L. J. Zhou, G. Rixecker, A. Zimmermann, and F. Aldinger, Electric fatigue in anti-ferroelectric Pb$_{0.97}$La$_{0.02}$(Zr$_{0.55}$Sn$_{0.33}$Ti$_{0.12}$)O$_3$ ceramics induced by bipolar cycling, J. Eur. Ceram. Soc., 26, 883-889 (2006).
14. X. Hao, J. Zhai, X. Yao, Improved energy storage performance and fatigue endurance of Sr-doped PbZrO$_3$ anti-ferroelectric thin films," J. Am. Ceram. Soc., 92, 1133–1135 (2009).
15. G. Zhang, D. Zhu, X. Zhang, L. Zhang, J. Yi, B. Xie, Y. Zeng, Q. Li, Q. Wang, S. Jiang, High-energy storage performance of (Pb$_{0.87}$Ba$_{0.1}$la$_{0.02}$)(Zr$_{0.68}$Sn$_{0.24}$Ti$_{0.08}$)O$_3$ antiferroelectric ceramics fabricated by the hot-press sintering method. J. Am. Ceram. Soc. *98*, 1175-1181 (2015).
16. G. L. Tan, W. Li, Ferroelectricity and ferromagnetism of M-type lead hexaferrite, J. Am. Ceram. Soc. **98**, 1812-1817 (2015).
17. Xue Li, Guo-Long Tan, Multiferroic and magnetoelectronic polarizations in BaFe$_{12}$O$_{19}$ system, Journal of Alloys and Compounds, 858, 157722/1-15 (2021).
18. G. L. Tan, Y. Huang, H. H. Sheng, Magnetoelectric Response in Multiferroic SrFe$_{12}$O$_{19}$ Ceramics, PLos One, 11, e0167084/1-21 (2016).
19. G. L. Tan, N. Nan, P. Sharma, A. Kumar, Antiferroelectric and magnetic performance in La$_{0.2}$Sr$_{0.7}$Fe$_{12}$O$_{19}$ system, J. Mater. Sci.: Mater Electron 32, 21697–21708 (2021).
20. I. Nagakura, T. Ishii, T. Sagawa, Photoelectron spectroscopic measurement of core level binding energies of Yttrium, Lanthanum and Cerium, J. Phys. Soc. Jap. 33, 754-760 (1972).
21. K. Ichikawqa, O. Aita, K. Aoki, Nonradiative decay processes of 4d hole states in CsF, BaF$_2$, LaF$_3$, Phys. Rev. B 45, 3221-3229 (1992).
22. H. R. Jo and C. S. Lynch, A high energy density relaxor antiferroelectric pulsed capacitor dielectric, J. Appl. Phys. **119**, 024104 /1-7 (2016).
23. Q. Zhang, H. Tong, J. Chen, Y. Lu, T. Yang, X. Yao, Y. He, High recoverable energy density over a wide temperature range in Sr modified (Pb, La)(Zr,Sn,Ti)O$_3$ antiferroelectric ceramics with an orthorhombic phase, Appl. Phys. Lett. 109, 262901/1-4 (2016).

24. L. Zhao, Q. Liu, J. Gao, S. Zhang, J. F. Li, Lead-free antiferroelectric silver niobate tantalite with high energy storage performance, Adv. Mater. **29**, 1701824/1-7 (2017).

25. T. Wang, J. Hu, H. Yang, L. Jin, X. Wei, C. Li, F. Yan, Y. Lin, J. Appl. Phys. **121**, 084103 (2017).

26. X. H. Hao, J. W. Zhai, L. B. Kong, Z. K. Xu, A comprehensive review on the progress of lead zirconate-based antiferroelectric materials, Progress in Materials Science 63, 1-57 (2014).

27. J. Parui, S. B. Krupanidhi, Enhancement of charge and energy storage in sol-gel derived pure and La modified $PbZrO_3$, Appl. Phys. Lett. 92, 192901/1-3 (2008).

28. L. T. Yang, X. Kong, F. Li, H. Hao, Z. X. Cheng, H. X. Liu, J. F. Li, S. J. Zhang, Perovskite lead-free dielectrics for energy storage applications, Progress in Materials Science 102, 72-108 (2019).

29. R. Chen, J. F. Wang, Z. W. Ouyang, Z. Z. He, S. M. Wang, L. Lin, J. M. Liu, C. L. Lu, Y. Liu, C. Dong, C. B. Liu, Z. C. Xia, A. Matsuo, Y. Kohama, K. Kindo, Magnetic field induced ferroelectricity and half magnetization plateau in polycrystalline $R_2V_2O_7$ ($R$ = Ni, Co), Phys. Rev. B 98, 184404/1-7 (2018).

30. Y. J. Choi, C. L. Zhang, N. Lee, S-W. Cheong, Cross-control of magnetization and polarization by electric and magnetic fields with competing multiferroic and weak-ferromagnetic phases, Phys. Rev. Lett., 105, 097201/1-4 (2010).

31. G. L. Tan, H. H. Sheng, Multiferroism and colossal magneto-capacitance effect of $La_{0.2}Pb_{0.7}Fe_{12}O_{19}$ ceramics, Acta Mater. 121, 144-151 (2016).

32. M. N. Baibich, J. M. Broto, A. Fert, F. Nguyen Van Dau, F. Petroff, P. Etienne, G. Creuzet, A. Friederich, J. Chazelas, Giant magnetoresistance of (001)Fe/(001) Cr magnetic supperlattices, Phys. Rev. Lett. **61**, 2472-2475 (1988).

33. G. Binash, P. Grünberg, F. Saurenbach, W. Zinn, Enahnced magnetoresistance in layered magnetic structures with antiferromagnetic interlayer exchange, Phys. Rev. B 39, 4828-4830 (1989).

34. A. Fert, P. Griinberg, A. Barthelemy, F. Petroff, W. Zinn, Layered magnetic structures: interlayer exchange coupling and giant magnetoresistance, J. Magn. Magn. Mater. 140-144, 1-8 (1995).

35. A. Fert, Nobel Lecture: Origin, development, and future of spintronics, Rev. Mod. Phys. 80, 1517-1530 (2008).
36. R. J. Tang, H. Zhou, W. L. You, H. Yang, Room temperature multiferroic and magnetocapacitance effects in M-type hexaferrite $BaFe_{10.2}Sc_{1.8}O_{19}$, Appl. Phys. Lett. 109, 082903/1-5 (2016).
37. Y. Feng, G. L. Tan, Magnetodielectric coupling response in La-modified M-type strontium hexaferrite, Phys. Status Solidi A 215, 1800295/1-7 (2018).